\documentclass[preprint,aps,prl,groupedaddress,amssymb]{revtex4}
\usepackage{graphicx}
\usepackage{dcolumn}
\usepackage{amsmath}
\begin{document}

\title{Dielectric photonic crystal as medium with negative electric permittivity and magnetic permeability}

\author{A. L. Efros}
\email{efros@physics.utah.edu}
\affiliation{University of Utah, Salt Lake City UT, 84112 USA}

\author{A. L. Pokrovsky}
\affiliation{University of Utah, Salt Lake City UT, 84112 USA}

\date{\today}

\begin{abstract}

We show that a two-dimensional photonic crystal (PC) made from a {\it non-magnetic} 
dielectric is a left-handed material in the sense defined by Veselago. 
Namely, it has negative values of both the electric permittivity $\epsilon$ and the 
magnetic permeability $\mu$ in some frequency range. 
This follows from a recently proven general theorem. 
The negative values of $\epsilon$ and $\mu$
are found by a numerical simulation.
Using these values we demonstrate the Veselago lens, 
a unique optical device predicted by Veselago. 
An approximate analytical theory is proposed to calculate the values of $\epsilon$ and $\mu$ from 
the PC band structure. 
It gives the results that are close to those obtained by the numerical simulation. 
The theory explains how a non-zero magnetization arises in a non-magnetic PC.

\end{abstract}

\maketitle

About 35 years ago Victor Veselago\cite{ve} considered theoretically 
propagation of the electromagnetic waves (EMW's) in a hypothetical left-handed medium (LHM), 
where both the electric permittivity $\epsilon$ 
and the magnetic permeability $\mu$ are negative in some frequency range. 
Since the square of the speed of light $c^2 =c_0^2/\epsilon \mu$ 
is positive, the EMW's propagate but they have unusual properties, 
like anomalous Doppler and Cherenkov effects, negative light pressure, 
and negative refraction (NR) at the interface with a regular medium. 
The latter property enables three-dimensional (3D) imaging. 
The origin of all these properties is the fact that the energy flux, {\bf S} 
and the wave vector, {\bf k} are antiparallel in the LHM. 
An important assumption of Veselago's theory is the absence of spatial
dispersion, which means that both $\epsilon$
and $\mu$ are ${\bf k}$-independent. However both of them are functions of frequency. 
A generalization of Veselago's theory for the case of spatial dispersion
would not be simple because in this case
the expression for the Poynting vector includes an extra term \cite{lan}. 

To create the LHM Veselago proposed to combine two independent 
different subsystems \cite{ve}, one 
with a negative $\epsilon$ and another with a negative $\mu$.
As an example, he considered a ferromagnetic metal at frequency close to the 
ferromagnetic resonance and below the plasma frequency of the metal. 
Following this line, the San Diego group\cite{sm3} has observed 
a NR in an artificial composite system and attributed this effect to 
a two-dimensional (2D) LHM. 
Their system consists of two elements: 2D metallic photonic crystal (PC) creating
negative $\epsilon$ below a cutoff frequency and the so-called split-ring resonators creating
negative $\mu$ due to anomalous dispersion near the resonance. 
The experiments of the 
San Diego group are very important because they have ignited the whole field, 
started by Veselago about 35 years ago but completely forgotten since then.
However they caused new hot debates concerning different 
aspects of their experiments and interpretations\cite{loww,val,smsk,smsch,nv1}.

Many groups\cite{kosaka,notomi,jj,jj3d,cub} 
predicted and observed NR in PC's using for the prediction the dispersion relation of 
EMW's and methods of geometrical optics. 
However, the ``effective refractive index'' they found was, in 
general, a function of ${\bf k}$. 
We think that it originates from the spatial dispersion of the dielectric constant for EMW 
with ${\bf k}$ deep in the Brillouin zone. 
Several authors\cite{kosaka,notomi,jj3d} found that near the $\Gamma$ point this index 
becomes $k$-independent. 
As we show below, together with the negative group velocity this indeed
is a manifestation of the LHM. 

The main idea of this work is based upon the theorem proved in Ref. \cite{n}.
It reads that the group velocity ${\bf v}_g = \partial \omega /\partial {\bf k}$
in an isotropic medium is positive in the regular media (RM)($\epsilon>0$, $\mu>0$)
and negative in the LHM. This means that vectors $\partial \omega /\partial {\bf k}$
and  ${\bf k}$ are parallel in the RM and antiparallel in the LHM.
This theorem follows from the thermodynamics requirement that 
the energy of electromagnetic fields is positive. 

\begin{figure}
\includegraphics[width=8.6cm]{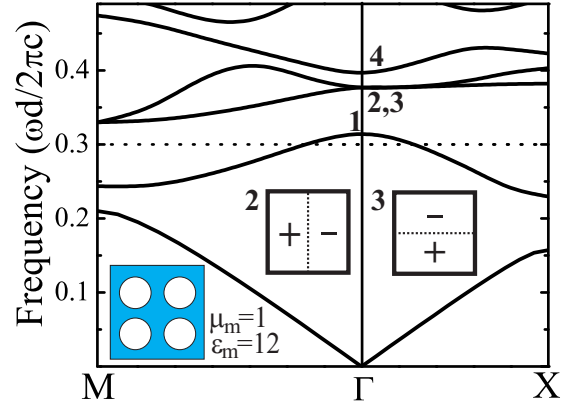}
\caption{The lowest bands of the 2D photonic crystal that is shown  
in the left bottom inset. 
The cylindrical hole radius is $0.35 d$. 
Dashed line indicates the working frequency. 
Central insets show signs of electric field in a unit cell for bands $2$, $3$ at $k=0$. 
Points $M$ and $X$ are at the boundary of the Brillouin zone in directions $[1,1]$ and $[1,0]$ respectively. \label{fig1} }
\end{figure}

Now we generalize this theorem to a 2D PC illustrated in the left 
bottom inset of Fig. \ref{fig1}. The PC represents 
a square lattice of period $d$, comprising cylindrical holes 
in a dielectric medium with the axes of cylinders along $z$-direction. 
We consider propagation of EMW's with the electric field ${\bf E}$ in this direction (s-wave). 
The calculated \cite{weusef} low frequency 
spectrum of EMW's in this PC is shown in 
Fig. \ref{fig1} for the case when the dielectric material of the crystal is 
characterized by $\epsilon_m=12$, $\mu_m=1$. 
The fields in the PC are represented by the Bloch functions with a 2D wave vector ${\bf k}$=($k_x$, $k_y$).
One can see from Fig. \ref{fig1} that in the frequency range a little below the 
top of the band $1$ at the $\Gamma$ point, the group velocity 
$\partial \omega /\partial {\bf k}$ is 
negative and there are no other bands in this frequency range. 
It is also important that band $1$ is not degenerate at $\Gamma$-point. 
Near the $\Gamma$-point the wavelength is much larger than the lattice 
period of the PC  
and one can introduce ${\bf k}$-independent $\epsilon_{ik}$ and $\mu_{ik}$. 
Since $z$-axis is the axis of $C_4$ symmetry, we have a uniaxial crystal and both 
tensors in the principal axes have only two components 
$\epsilon_{zz}$, $\epsilon_{\perp}= \epsilon_{xx}= \epsilon_{yy}$ and 
$\mu_{zz}$, $\mu_{\perp} = \mu_{xx} = \mu_{yy}$ respectively. 
The electromagnetic energy density of the s-wave can be 
represented in the form\cite{lan}
\begin{equation}
\label{u}
\bar{U} = \frac{1}{16 \pi \mu_{\perp} \omega} \frac{d[\omega^2 n^2]}{d \omega} |E|^2,
\end{equation}
where $n^2=\epsilon_{zz} \mu_{\perp}$. 
In the rest of the paper we use notation 
$\epsilon_{zz} = \epsilon$, $\mu_{\perp} = \mu$.
Since $\bar{U} >0$ we get
$\mu({\bf v}_g \cdot {\bf k})>0$, where 2D group velocity 
${\bf v}_g =2c_0^2 {\bf k}/[d(\omega^2 n^2)/d \omega]$. 
Thus, negative ${\bf v}_g \cdot {\bf k}$ means negative $\mu$ and vise versa. 
For the propagating waves it also means $\epsilon < 0$.  
It follows that this PC is a LHM in the frequency range under study.

Note, that the p-wave (magnetic field in $z$ direction) is characterized by 
$\epsilon_{\perp}$ and $\mu_{zz}$. 
These components might also be negative in the vicinity of the $\Gamma$ point 
but within a different frequency range.

\begin{figure}
\includegraphics*[width=16.0cm]{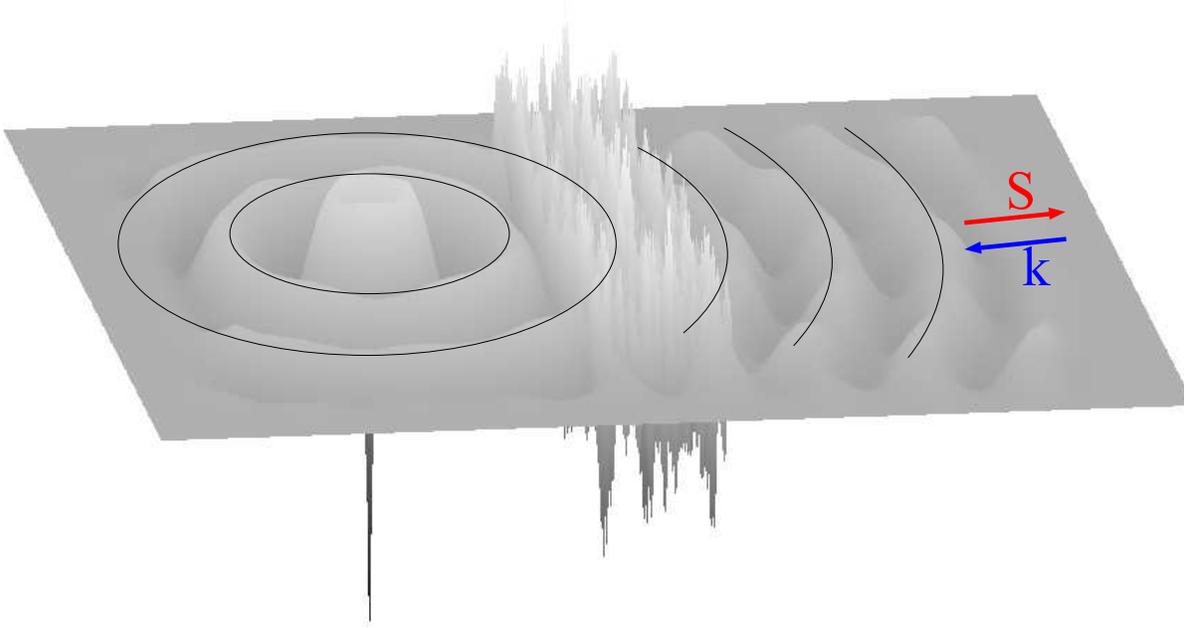}
\caption{Finding the values of $\epsilon$ and $\mu$ of PC using the propagation 
of EMW's from the 2D point source through the slab of the PC. 
Parameters of the homogeneous medium are $\epsilon' =-1.2$, $\mu' =-0.096$. 
Black curves show the circles centered at the point source. 
One can see that EMW's in the homogeneous medium almost ignore the slab. 
We conclude from this that $\epsilon = \epsilon'$, $\mu = \mu'$. 
Vectors ${\bf S}$ and ${\bf k}$ are opposite to each other as it should be in the LHM. 
The computations are performed by the finite-element method with perfectly matched layers at the boundaries.
\label{fig2} }
\end{figure}

We find $\epsilon$ and $\mu$ by microscopic calculations and show that they are
indeed both negative in the frequency range under study. 
For this purpose we consider a PC slab infinite in $z$-direction and embedded 
in a homogeneous medium with $\epsilon'$, $\mu'$. 
A 2D point source (an infinite dipole along $z$-axis) is located in the homogeneous medium 
to the left of the slab. 
If we now choose $\epsilon'$ and $\mu'$ so that the cylindrical 
wave in the system is the same as it would be in a
completely homogeneous medium with $\epsilon'$ and $\mu'$ (though inside 
the slab there is a microscopic field), the values of $\epsilon'$ and $\mu'$ obtained in this way are the 
macroscopic $\epsilon$ and $\mu$ of the slab at a given frequency. 
The best choice for $\epsilon'$ and $\mu'$ at the working 
frequency $\omega = 0.3 (2 \pi c_0/d)$ 
is obtained by a fitting procedure and the result 
is shown in Fig. \ref{fig2}. 
It yields $\epsilon=-1.2$, $\mu=-0.096$. 
Thus, at this frequency the PC slab is a LHM  and this result is in accordance with the above theorem. 
Note, that in the case of the LHM the vectors {\bf S} and {\bf k} are in the opposite directions 
(see animation \cite{ourweb}).

One can calculate independently the refractive 
index of the slab $n^2=\epsilon \mu$ 
from the dispersion relation of 
the band $1$. At small $k$ it can be written in a form 
$\omega^2=\omega_1^2-\alpha c_0^2 k^2$.
Our calculation of spectrum 
gives $\alpha=0.94$, $\omega_1 = 0.314 (2 \pi c_0/d)$. 
Using the alternative definition of $n^2$ namely, $\omega^2 n^2 = c_0^2 k^2$ 
one gets 
\begin{equation}
\label{n2}
n^2=\frac{1}{\alpha}\left(-1+ \frac{\omega_1^2}{\omega^2}\right). 
\end{equation}
Note, that $n \rightarrow 0$ as $\omega \rightarrow \omega_1$.
At the working frequency $n^2 = (0.32)^2$, which agrees with $\epsilon$ and $\mu$ found above.

\begin{figure}
\includegraphics*[width=8.6cm]{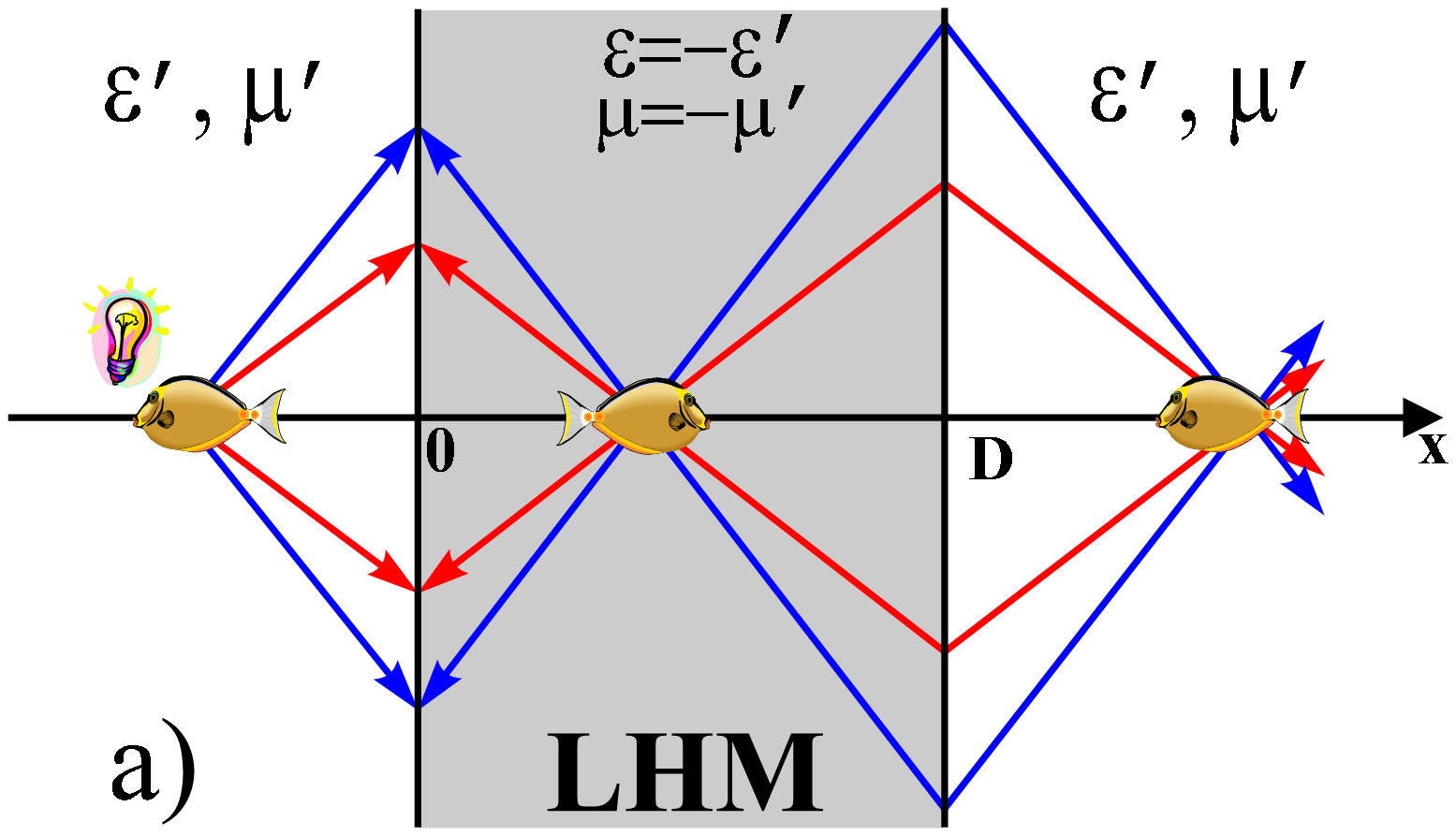}
\includegraphics*[width=8.6cm]{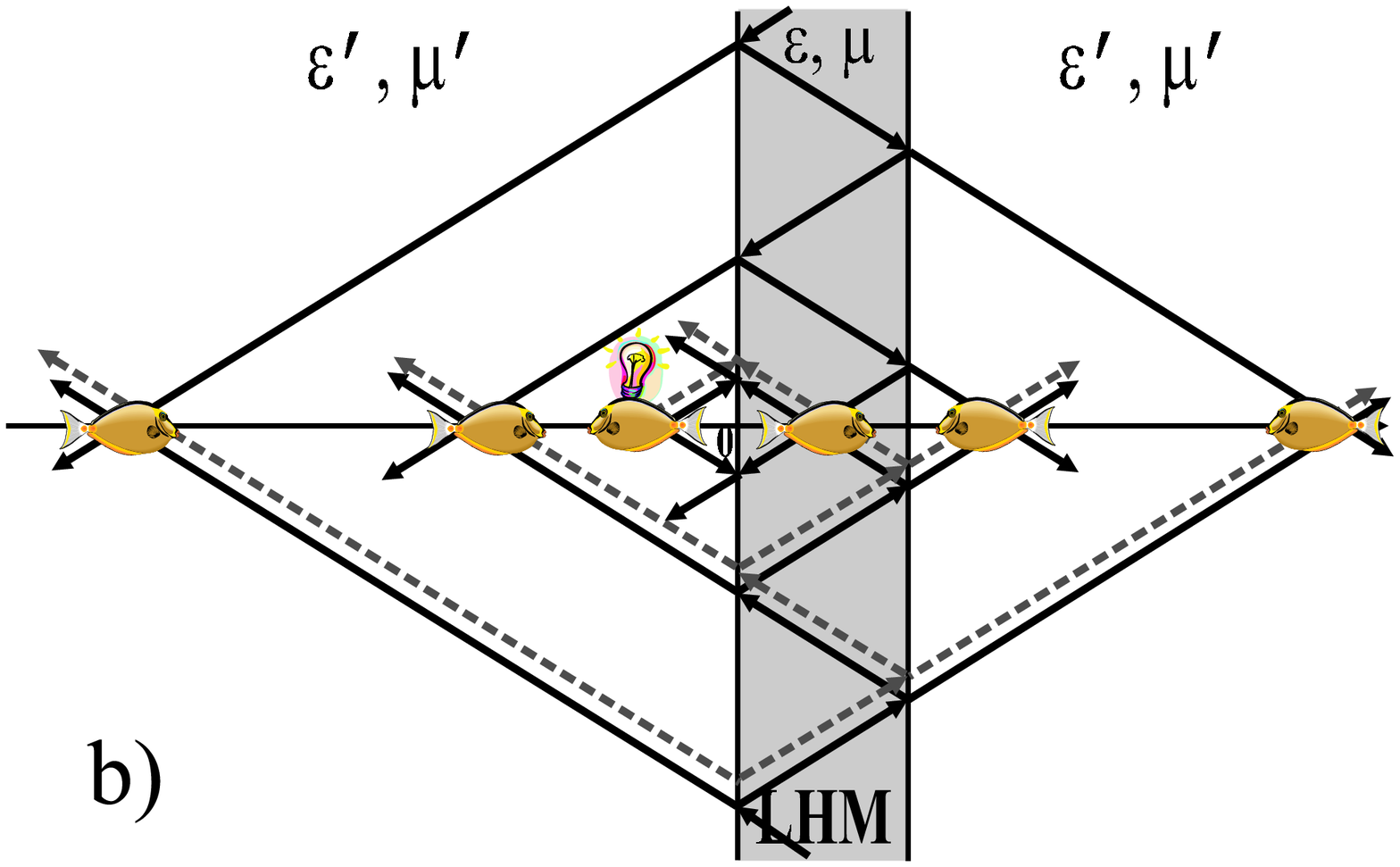}
\caption{ 3D imaging by the lenses based upon LHM. 
The objects are marked by bulbs.
The negative refraction at the interfaces is obtained from 
electrodynamics conditions of continuity of the tangential component 
of ${\bf k}$ and the normal component of ${\bf S}$, 
and from the main LHM property that the direction of 
${\bf S}$ is opposite to the direction of ${\bf k}$ (shown by arrows). 
a) The lens proposed by Veselago (Veselago lens\protect\cite{ve}). 
In this case $\epsilon' = |\epsilon|$, $\mu' = |\mu|$, 
the reflected waves are absent and the lens has only one external focus.
b) Multifocal lens\protect\cite{newlens} with $\epsilon \mu = \epsilon' \mu'$. The multiple foci
appear due to reflected waves. \label{fig3} }
\end{figure}

\begin{figure}
\includegraphics*[width=16.0cm]{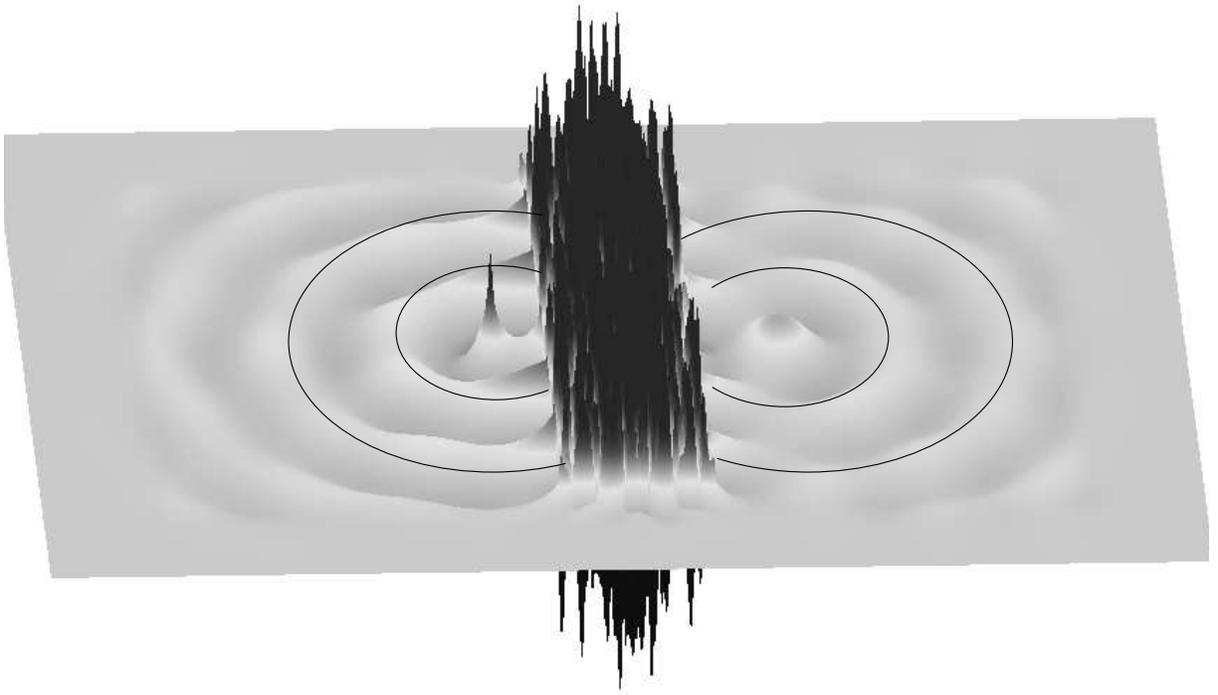}
\caption{Computer simulation of the 2D Veselago lens. 
In this case $\epsilon' =|\epsilon|$, $\mu'=|\mu|$. 
The curves show the circles centered in the point source to the left 
of the PC slab and at the focus to the right of the slab. 
Details of the simulation are similar to Fig. 2.\label{fig4} }
\end{figure}

If we choose $\epsilon'=|\epsilon|$ and $\mu'=|\mu|$ as the parameters of the 
homogeneous medium surrounding the PC, we obtain a unique device proposed by Veselago\cite{ve} 
that we call the Veselago lens. 
It is shown schematically in Fig. \ref{fig3}a. 
The computer simulation of the Veselago lens is shown in Fig. \ref{fig4} 
(also see animation \cite{ourweb}). 
In the above simulations the PC is cut in the direction $[10]$. 
We have checked that the results do not change if the PC is cut in the direction $[11]$. 
This confirms the absence of spatial dispersion at the working frequency. 
The size of the focus is of the order of the wavelength as predicted by the 
diffraction theory \cite{focus}.

Now we explain how the starting theorem that follows from Eq.(\ref{u}) and 
is formulated in terms of macroscopic 
electrodynamics can be interpreted microscopically. 
The negative group velocity shown at Fig. \ref{fig1} 
results from the repulsion of bands $1$ and $2$, $3$.  
Therefore a microscopic explanation of negative $\mu$ should 
be found in terms of the interaction between these bands. 
We propose a quantitative explanation by writing Maxwell's equation for 
the microscopic field $e_z({\bf r})$ in a matrix form\cite{joan_pc,sak} 
in the basis of Bloch functions at $k=0$. 
Taking into account that near the $\Gamma$-point the bands $1$, $2$, $3$, $4$ 
are far from the other bands, we truncate the matrix keeping only these four bands. 
This approximation is analogous to the Kane model in semiconductor physics\cite{kane}.
By diagonalizing the truncated matrix we find the 
spectra of all four bands and the 
correct combinations of basis 
functions for each band at finite $k$. 
We have checked that the dispersion law for all bands $1$-$4$ as obtained
from the truncated model does not differ by more than $4\%$ from the result presented in Fig. \ref{fig1} 
at $kd/2\pi < 0.1$.
At small $k$ the microscopic electric field for the band 
$1$ has the form
\begin{equation}
\label{ez}
e_z({\bf r}) = e^{i {\bf k} \cdot {\bf r}}\left(
u_1 - \frac{i k_0 k_x c_0^2}{\omega_2^2-\omega_1^2}u_2 - \frac{i k_0 k_y c_0^2}{\omega_2^2-\omega_1^2}u_3-
\frac{i k_0 k_1 k^2 c_0^4}{(\omega_2^2-\omega_1^2)(\omega_4^2-\omega_1^2)}u_4 \right),
\end{equation}
where ${\bf r}$ is the 2D-vector, 
$u_1$, $u_2$, $u_3$ and $u_4$ are the Bloch functions at $k=0$, the 
positive parameters $k_0$, $k_1$ have the dimensionality of a wave vector and they are 
integrals from these functions.
The symmetry of the function $u_1$ is $X^2-Y^2$. The function $u_4$ has $X^2+Y^2$ symmetry.
The functions $u_2$ and $u_3$ have the symmetries $X$ and $Y$ respectively. 
The parity of $X$ and $Y$ functions within the elementary cell 
is shown in the central inset of Fig. \ref{fig1}.

If the material of the PC were metallic, the electric field in the 
bands $2$ and $3$ 
would create two currents in opposite directions. 
These currents would induce a magnetic moment in $x$ and $y$ directions. 
In case of the dielectric PC these currents are imaginary displacement 
currents and they do not create any magnetic moment in the modes $2$ and $3$. 
However, as follows from Eq.(\ref{ez}) the admixture 
of bands $2$ and $3$ to band $1$ is also imaginary. 
Therefore the band $1$ has a real magnetic moment.

The magnetization can be calculated using the expression
\begin{equation}
{\bf M} = \frac{1}{2 c_0 d^2} \int {\bf r} \times \frac{\partial {\bf P}}{\partial t} ds,
\end{equation}
where ${\bf P} = {\bf z} e_z({\bf r}) [\epsilon({\bf r})-1]/4\pi$, ${\bf z}$ is a unit vector
in $z$ direction, $\epsilon({\bf r})= 1$ inside the holes and $\epsilon_m$ otherwise, 
the integration is performed over the unit cell, 
and $e_z({\bf r})$ is given by Eq.(\ref{ez}). 
Only the second and the third terms from Eq.(\ref{ez}) contribute to the magnetization. 
The result can be expressed through the integrals from $u_2$ and $u_3$. 
These functions 
have been taken from the computation of the spectrum shown at Fig. \ref{fig1}.
The macroscopic vectors ${\bf B}$ and ${\bf E}$ are connected by the relation 
${\bf B} = c_0 {\bf k} \times {\bf E}/\omega$, where ${\bf E} = {\bf z} <e_z({\bf r})>$ 
and $<...>$ means averaging over the unit cell. 
Only the function $u_4$ gives a non-zero contribution after the averaging.
It gives the main contribution to ${\bf E}$. Other basis functions give non-zero
contribution 
only when multiplied by ${\bf r}$-dependent terms in the expansion of the 
 $\exp{(i{\bf k}\cdot {\bf r})}$ within the unit cell. 
Their contributions to ${\bf E}$ are of the same order in $k$ but they are much smaller
than the contribution from $u_4$ since $(\omega_4-\omega_1)/\omega_1 \ll 1$, 
where $\omega_1$ and 
$\omega_4$ are the frequencies of the bands $1$ and $4$ at $k=0$.
In fact, the latter strong inequality permits truncation of the matrix in our analytical model.

Now we can find $\mu$ from the definition $\mu^{-1}=1-4\pi|{\bf M}|/|{\bf B}|$. 
The result is 
\begin{equation}
\label{mu}
\mu=0.89 (1-\omega_1^2/\omega^2).
\end{equation} 
Comparing Eq.(\ref{n2}) and Eq.(\ref{mu}) one can see that the root of $\mu$ 
at $\omega =\omega_1$ is responsible for the root of $n^2$. 
This means that $\epsilon$ is non-zero and negative at $\omega = \omega_1$.
On the other hand, $\mu$ changes its sign at $\omega = \omega_1$. Thus, the gap at $\omega > \omega_1$
is due to $\mu \epsilon < 0$. 
At $\omega=0.3 (2 \pi c_0/d)$ one gets $\mu=-0.084$ and that is very close to the value 
obtained above by the fitting procedure. 
Thus, we have found analytically the value of $\mu$.

The 3D LHM can be obtained in the same way, for example, 
using the PC with the cubic symmetry proposed by Luo {\it et al.}\cite{jj3d}. 
This PC also has a negative group velocity near the $\Gamma$ point with 
no other bands at this frequency. 

Thus, we have shown that the LHM can be readily obtained  using 
only one 
dielectric PC that creates both negative 
$\epsilon$ and $\mu$. We have presented 
the physical explanation of this phenomenon in a 2D case.  
The working frequency can be easily changed in a wide range 
by changing the lattice constant and $\epsilon_m$ of the PC.  
It is important for applications 
that the losses in a dielectric PC are much smaller than in a metallic one. 
We have found not only $n$ but also $\epsilon$ and $\mu$ 
that are necessary to describe the dynamical properties of the interfaces. 
This is especially important for the creation of the 
Veselago lens, which requires no reflection at interfaces, thus imposing 
two matching conditions on $\epsilon$ and $\mu$. 
By matching $n$ only one gets a modification of the 
Veselago lens with multiple foci (Fig. \ref{fig3}b)\cite{newlens}.
We realize, that we have not presented a method of matching the LHM with the surrounding medium.
This matching is very easy in computational approach but it must be handled 
creatively on the way toward real devices.

\begin{acknowledgments}
We are grateful to Serge Luryi and Valy Vardeny
for helpful discussions.
The work has been funded by the NSF grant DMR-0102964. 
\end{acknowledgments}

\bibliography{neg}

\begin{thebibliography}{21}
\bibitem[{weu()}]{weusef}
\bibinfo{note}{All computations in this paper are performed using the
  commercially available software FEMLAB.}
\end{thebibliography}

\end{document}